\documentclass[10pt,twocolumn,conference]{IEEEtran}
\usepackage{amsmath, graphics,amssymb,epsfig,subfigure}
\usepackage{algorithm}
\usepackage{algorithmic}
\usepackage{float}
\usepackage{multirow}
\usepackage{cuted,flushend}
\usepackage{midfloat}
\begin{document}
\title{{Signal Space Alignment for an Encryption Message and Successive Network Code Decoding on the MIMO $K$-way Relay Channel} }


\author{{Namyoon Lee}$^{\dagger}$
        and {Joohwan Chun}$^{\ddagger}$ \\
\authorblockA{${}^{\dagger}$Samsung Advanced Institute of Technology (SAIT), Yongin-si, Korea \\
${}^{\ddagger}$Electrical Engineering department, KAIST, Daejeon, Korea \\
E-mail: namyoon.lee@gmail.com, and chun@ee.kaist.ac.kr}}

\maketitle

\begin{abstract}
This paper investigates a network information flow problem for a
multiple-input multiple-output (MIMO) Gaussian wireless network
with $K$-users and a single intermediate relay having $M$
antennas. In this network, each user intends to convey a multicast
message to all other users while receiving $K-1$ independent
messages from the other users via an intermediate relay. This
network information flow is termed a MIMO Gaussian $K$-way relay
channel. For this channel, we show that $\frac{K}{2}$ degrees of
freedom is achievable if $M=K-1$. To demonstrate this, we come up
with an encoding and decoding strategy inspired from cryptography
theory. The proposed encoding and decoding strategy involves a
\textit{signal space alignment for an encryption message} for the
multiple access phase (MAC) and \textit{zero forcing with
successive network code decoding} for the broadcast (BC) phase.
The idea of the \emph{signal space alignment for an encryption
message} is that all users cooperatively choose the precoding
vectors to transmit the message so that the relay can receive a
proper encryption message with a special structure,
\textit{network code chain structure}. During the BC phase,
\emph{zero forcing combined with successive network code decoding}
enables all users to decipher the encryption message from the
relay despite the fact that they all have different
self-information which they use as a key.
\end{abstract}


\section{Introduction}
Two-way communication or bidirectional communication between two
nodes was first considered by Shannon \cite{C. E. Shannon}. By
involving a relay node between two nodes, the two-way relay
channel has drawn great interest from numerous researchers due to
its useful application scenarios in cellular and ad-hoc networks.
Several studies \cite{Rankov1}-\cite{Zhang2} showed that the
sum-rate performance of the two-way relaying protocol
significantly increases compared with that of one-way relay
protocol.

Recently, the two-way relay channel has been generalized in
various network information flow settings. To realize multi-pairs
information exchange, the authors in \cite{Chen}-\cite{Sezgin}
studied a multi-pair two-way relay channel in which the relay
helps the communication between multiple pairs of users. In
\cite{Chen}, authors proposed a jointly demodulate-and-XOR forward
relaying scheme for a code-division multiple access system under
an interference-limited environment. In addition, other authors
\cite{Avestimehr}-\cite{Sezgin} characterized the capacity of a
multi-pair relay channel in a deterministic channel model. They
showed that a divide-and-conquer strategy exploiting signal-level
alignment for the MAC phase and a simple equation-forwarding
scheme for the broadcast phase (BC) achieve the capacity for a
deterministic channel case. They also showed these schemes
extended to a Gaussian channel case for achieving the capacity
within a constant number of bits.

\begin{figure}
\centering
\includegraphics[width=3.5in]{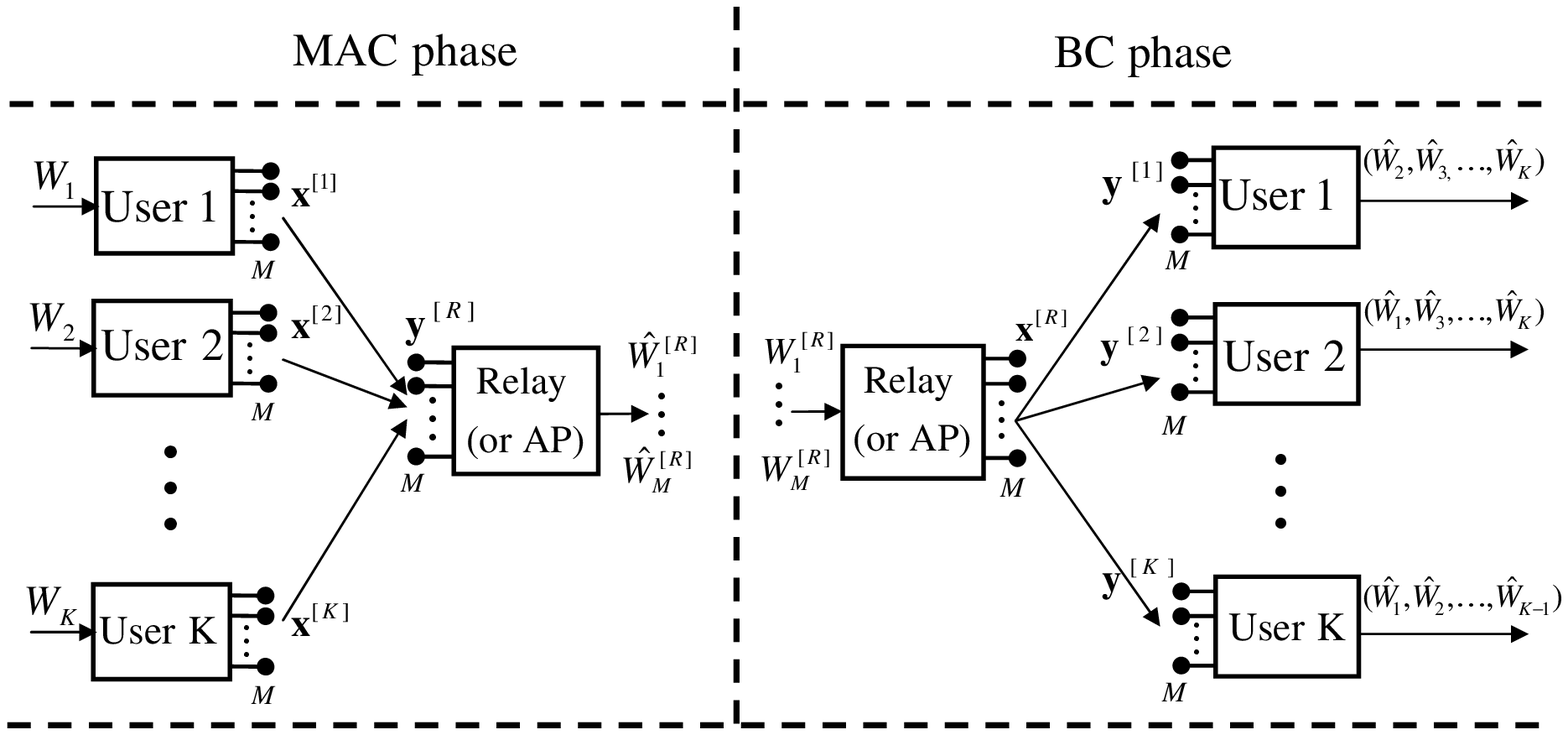}
\caption{System model of MIMO Gaussian $K$-way relay channel.}
\label{fig_sim}
\end{figure}

Furthermore, the authors in \cite{Gunduz2}-\cite{namyoon2}
investigated multiuser and multi-way relay communications, which
are more general scenarios than the use of a multi-pair two-way
relay channel in the terms of multiple directions of
communication. In \cite{Gunduz2}, for several relaying schemes,
examined the achievable rate regions of a multi-way relay channel
with multiple clusters of users such that the users in each
cluster want to exchange multicast message within that group of
users. By considering multiple unicast messages per user, in our
previous works \cite{namyoon1}-\cite{namyoon2}, an efficient
coding scheme termed the \textit{signal space alignment for
network coding} was proposed for a multiple-input-multiple-output
(MIMO) Gaussian Y channel. Using this scheme in \cite{namyoon2},
the authors showed that the degrees of freedom for the MIMO
Gaussian Y channel is $3M\log(\textsf{SNR})+o(\log(\textsf{SNR}))$
when each user has $M$ antennas and when the relay has $N \geq
\lceil{\frac{3M}{2}}\rceil$ antennas. From this scheme, the
authors demonstrated how to deal with multiple interference
signals on multi-user and multi-way communications by
appropriately exploiting interference alignment \cite{Cadambe} and
network coding \cite{Katti}-\cite{Zhang2}.

In this paper we investigate a network information flow problem
for a MIMO Gaussian wireless network with $K$-users and a single
intermediate relay having $M$ antennas i.e., the single cluster
multi-way relay channel in one of the previous studies
\cite{Gunduz2}. In this network, user $i$, $i=\{1,2,\ldots,K\}$
wants to convey a multicast message $W_i$ to all other users,
while receiving $K-1$ independent messages from the other users
via the intermediate relay. This network information flow is
termed a MIMO Gaussian $K$-way relay channel. There are various
interesting application scenarios of the MIMO Gaussian $K$-way
relay channel, such as video conferencing using Wi-Fi access point
and multi-player gaming using smart phones.

In this paper, we propose an encoding and decoding strategy which
were inspired by cryptography theory. In cryptography, encryption
is a process of transforming information using an algorithm to
make it readable to only those with special knowledge; this is
referred to as a key. Using this cryptography concept, the
conventional two-way relay channel can be interpreted from a new
angle. In a two-way relay channel, during the MAC phase, user 1
and user 2 transmit message $W_1$ and $W_2$ to the relay, and the
relay jointly detects and decodes the modulo sum of these two
messages, i.e., $W_1\oplus W_2$. This process can be thought as a
type of encryption algorithm. During the BC phase, each user can
decipher the encrypted message $W_1\oplus W_2$ from the relay
using its self-information as a key. Therefore, our encoding and
decoding problem in the MIMO Gaussian $K$-way relay channel is to
find the encryption algorithm so that all $K$ users who take part
in an information exchange can decode information from other users
using their own self-information as a key.

The newly proposed encoding and decoding strategy involves a
\textit{signal space alignment for an encryption message} for the
MAC phase and a \textit{zero-forcing combined with successive
network code decoding} for the BC phase. The key idea of the
\emph{signal space alignment for an encryption message} is that
all users cooperatively design precoding vectors for transmitting
a message so that the relay can receive \emph{a proper encryption
message}. Here, a proper encryption message is an encrypted
message that all users who take part in the information exchange
can resolve even if they all have different keys. In order for the
relay to have the properly encrypted message, the each user
selects the transmit signal direction for the encrypted
information so that it has a special structure, \textit{network
code chain structure}. During the BC phase, \emph{zero-forcing
combined with successive network code decoding} enables all users
to decipher the encryption message from the relay, even if they
all have different self-information. Because the encryption
message having a network code chain structure can be untangled
successively at all users' side by using own side-information.
From this decoding process, each user obtains the messages from
the other users on the network.

The organization of this paper is followings: Section
\ref{sec:Systemmodel} describes the system model of the MIMO
Gaussian $K$-way relay channel. In Section \ref{sec:Gaussian
channel}, we compare the achievable degrees of freedom of the MIMO
Gaussian $K$-way relay channel according to various schemes.
Finally, Section \ref{sec:Conclusions} concludes this paper.

\textrm{\textbf{Notation}}: We use bold upper and lower case
letters for matrices and column vectors, respectively.
$(\cdot)^{T}$ and $(\cdot)^{H}$ represent a transpose and a
Hermitian transpose, respectively. $\mathbb{E}(\cdot)$ and
$\textrm{Tr}({\bf A})$ denote the expectation operator and trace
of the matrix ${\bf A}$, respectively.

\section{System model}\label{sec:Systemmodel}
The MIMO Gaussian $K$-way relay channel shown in Fig. 1 is
considered in this section. In this channel, $K$ users and a relay
have $M$ multiple antennas. The users want to exchange messages
each other with the help of a single relay terminal. User $i$
wants to send message $W_i$ to the other users on the network and
intends to decode all other users' messages on the network, i.e,
$\{\hat{W_1},\hat{ W_2},\ldots, \hat{W_K}\}/ \{ \hat{W_i}\}$.

In multiple access (MAC) phase (the first time slot), user $i$
sends message $W_i$ to the relay. The received signal at the relay
is represented by
\begin{eqnarray}
{\bf y}^{[R]}&=&\sum_{i=1}^{K}{\bf H}^{[R,i]}{\bf x}^{[i]}+{\bf
n}^{[R]},
\end{eqnarray}
where ${\bf H}^{[R,i]}$ represents the $M\times M$ channel matrix
from user $i$ to the relay, ${\bf x}^{[i]}\in \mathbb{C}^{M}$
denotes transmit vector at user $i$, and ${\bf n}^{[R]}\in
\mathbb{C}^{M}$ denotes an additive white Gaussian noise (AWGN)
vector. The user has an average power constraint,
$\mathbb{E}\left[\textrm{Tr}\left({\bf x}^{[i]}{\bf
x}^{[i]^{H}}\right)\right]\leq \textsf{SNR}$. The channel is
assumed to be quasi-static and each entry of the channel matrix is
an independently and identically distributed (i.i.d.) zero mean
complex Gaussian random variable with unit variance, i.e.,
$\mathcal{NC}(0,1)$.

After receiving, the relay generates new transmitting signals and
broadcasts them to all users in what is known as, a BC phase. The
received signal vector at user $i$ is given by:
\begin{eqnarray}
{\bf y}^{[i]}&=&{\bf H}^{[i,R]}{\bf x}^{[R]}+{\bf n}^{[i]},
\end{eqnarray}
where ${\bf H}^{[i,R]}$ denotes the $M\times M$ channel matrix
from the relay to user $i$, ${\bf x}^{[R]}\in \mathbb{C}^{M}$ is
the transmit vector at the relay, and ${\bf n}^{[i]}\in
\mathbb{C}^{M}$ denotes the AWGN vector. The transmit signal at
the relay is subject to the average power constraint
$\mathbb{E}\left[\textrm{Tr}\left({\bf x}^{[R]}{\bf
x}^{[R]^{H}}\right)\right]\leq \textsf{SNR}$. If the system is
assumed to be in time-division-duplex (TDD) mode, the channel
${\bf H}^{[R,i]}$ is identical to ${\bf H}^{[i,R]^{H}}$. However,
it is assumed here that these two channels are generally different
by taking into account frequency-division-duplex (FDD). Throughout
this paper, direct links between users are neglected for
simplicity; thus, the relay is essential for communication.
Additionally, it is assumed that all users and the relay operate
in half-duplex mode. This implies that all terminals can not
receive and transmit simultaneously. The channel is assumed to be
known perfectly at all users and the relay in both the transmit
and receive modes.

\subsection{Multicast capacity and degrees of freedom }
By definition of multicast capacity in \cite{Jindal}, we define
the achievable rate for the multicast message $W_i$, as
$R_i(\textsf{SNR})$, which is given by
\begin{eqnarray}
R_i(\textsf{SNR}) = \min_{ j=\{1,2,\ldots,K\}/\{i\}}
R_{ji}(\textsf{SNR}),
\end{eqnarray}
where $R_{ji}(\textsf{SNR})$ is the achievable rate from user $i$
to user $j$, which is
\begin{eqnarray}
R_{ji}(\textsf{SNR}) = \min \left\{I\left({\bf x}^{[i]};{\bf
y}^{[R]}\right), I\left({\bf x}^{[R]};{\bf
y}^{[j]}\right)\right\}.
\end{eqnarray}
Therefore, the capacity region for the MIMO Gaussian $K$-way relay
channel can be defined as the set of all achievable rate tuples
$\mathcal{\bf
R}(\textsf{SNR})=\left[R_1(\textsf{SNR}),R_2(\textsf{SNR}),\ldots,R_K(\textsf{SNR})\right]^{T}$.


Furthermore, the degrees of freedom region for the MIMO Gaussian
$K$-way relay channel is defined as in (5) (Please see the top of
next page).
\begin{figure*}
    \begin{eqnarray}
    D^{Kw}\equiv \left\{{\bf d}\in \mathbb{R}^{K}_{+}:\forall {\bf
    w}\in \mathbb{R}^{K}_{+}\quad {\bf w}^{T}{\bf d}\leq
    \lim_{\textsf{SNR}\rightarrow
    \infty}\left[\sup_{\mathcal{R}(\textsf{SNR})\in C^{Kw}}\frac{{\bf
    w}^{T}\mathcal{\bf R}(\textsf{SNR})}{\log(\textsf{SNR})}\right]
    \right\}.\\ \hline \nonumber
    \end{eqnarray}
\end{figure*}
In (5), ${\bf d}=(d_1,d_2,\ldots,d_K)^{T}$, and ${\bf
w}=(w_1,w_2,\ldots,w_K)^{T}$. Accordingly, the sum of the degrees
of freedom $\eta$ is defined as
\begin{eqnarray}
\eta \triangleq \max_{D^{Kw}}(d_1+d_2+\ldots+d_K).
\end{eqnarray}

\section{Achievable degrees of freedom}\label{sec:Gaussian
channel}

In this section, we investigate the degrees of freedom that can be
achieved according to various schemes for the MIMO Gaussian
$K$-way relay channel. From this, we show that the proposed
encoding and decoding scheme is able to attain the higher degrees
of freedom in the MIMO Gaussian $K$-way relay channel by comparing
with that achieved by a conventional time-division-multiple-access
(TDMA) scheme.

\begin{figure}
\centering
\includegraphics[width=3.6in]{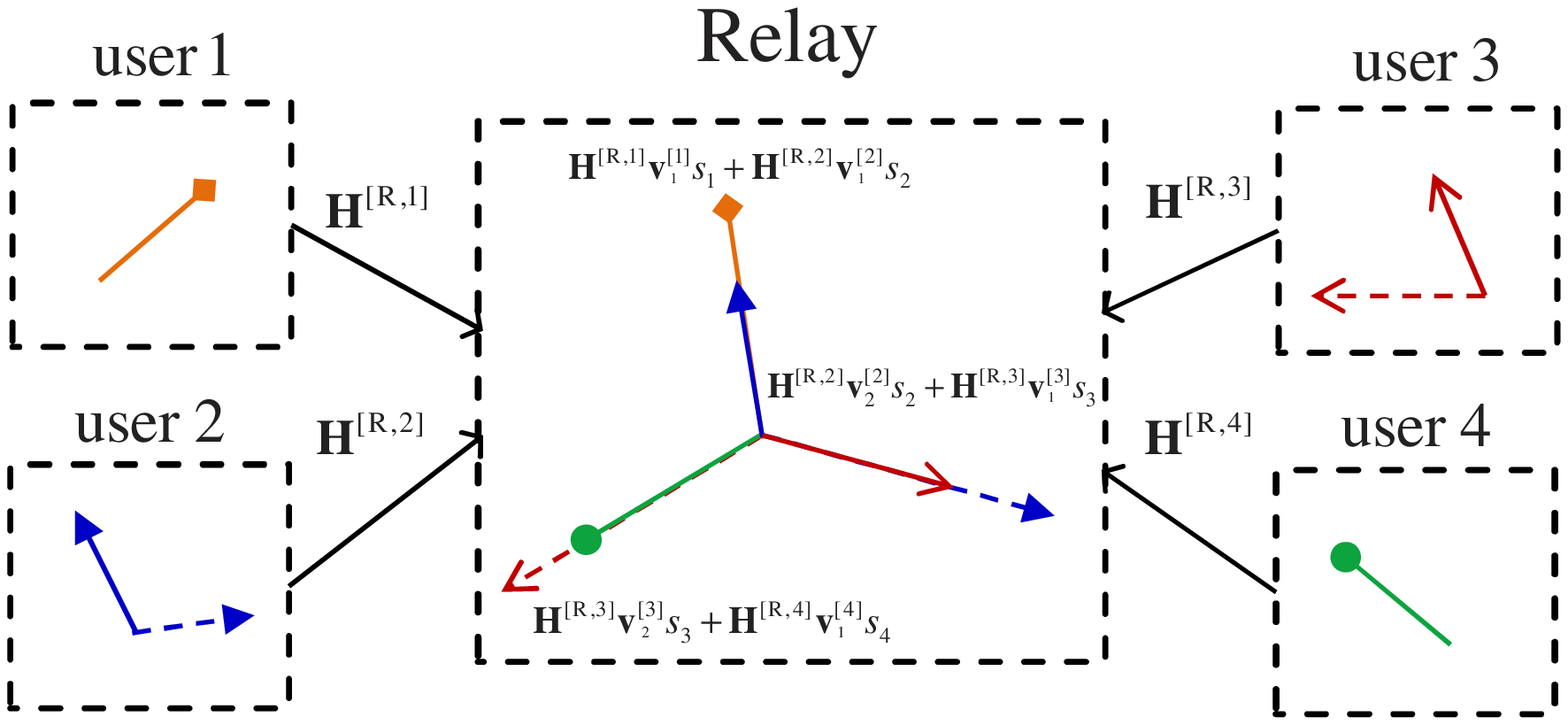}
\caption{Signal space alignment for encryption message for $K=4$
case.} \label{fig_sim}
\end{figure}
\subsection{TDMA scheme}
In the MIMO Gaussian $K$-way relay channel, TDMA scheme requires
$2K$ orthogonal time slots for completely exchanging the messages
each other via a relay. In the first time slot, user 1 transmits
the message $W_{1}$ using $M$ independent streams to the relay.
Thereafter, during the second time slot, the relay broadcasts the
received message using $M$ independent streams to all users
(multicasting). By applying zero-forcing spatial decoder, each
user can obtain the $M$ independent streams and decodes them to
get message, $W_1$. In a similar way, user $i$, $\{i=2,\ldots,K\}$
can deliver the message to the other users via the relay by
spending two orthogonal time slots. Therefore, the degrees of
freedom that can be achieved TDMA scheme for the MIMO Gaussian
$K$-way relay channel is given by:
\begin{eqnarray}
\eta_{TDMA} = \frac{KM}{2K}=\frac{K-1}{2}, \quad \textrm{if} \quad
M=K-1.
\end{eqnarray}

\subsection{Proposed scheme}
The following theorem is the main result of this paper.

\emph{\textbf{Theorem 1}}: In a MIMO Gaussian $K$-way relay
channel where each user and the relay have $M=K-1$ antennas,
$\frac{K}{2}$ degrees of freedom is achievable for $K\geq 2$.

Proof: Here, we only show the case of $K=4$ due to limitation of
space. The journal version of this paper \cite{namyoon3} includes
the proof for the general $K$-user case and it shows that the
proposed scheme can achieve the trivial outer bound derived by a
cut-set theorem in \cite{Cover}. The achievability proof is
provided using signal space alignment for an encryption message
and zero-forcing with successive network code decoding.

Using this encoding and decoding scheme, it is shown that
$(d_1,d_2,d_3,d_4)=(1,1,1,1)$, i.e.,
$\eta=\frac{\sum_{i=1}^{K}d_i}{2}=2$ is achieved.

For the MAC phase (the first time slot), user $1$ and user $4$
send messages $W_{1}$ and $W_{4}$ using symbols $s_1$ and $s_4$
along with precoding vectors ${\bf v}^{[1]}_1$ and ${\bf
v}^{[4]}_1$, as expressed by
\begin{eqnarray}
{\bf x}^{[1]}={\bf v}^{[1]}_1 s_{1}, \quad {\bf x}^{[4]}={\bf
v}^{[4]}_1 s_{4},
\end{eqnarray}
while user $2$ and user $3$ transmit message $W_2$ and $W_3$ using
symbol $s_2$ and $s_3$. Here, we assume
$\mathbb{E}\left[\|s_i\|^2\right]=1$. Contrast with $s_1$ and
$s_4$, $s_2$ and $s_3$ are transmitted two times along with two
different precoding vectors ${\bf v}^{[i]}_1$ and ${\bf
v}^{[i]}_2$, as expressed by
\begin{eqnarray}
{\bf x}^{[i]}={\bf v}^{[i]}_1 s_{i}+{\bf v}^{[i]}_2 s_{i}, \quad
i=\{2,3\}.
\end{eqnarray}

The received signal at the relay is given by
\begin{eqnarray}
{\bf y}^{[R]}=\sum_{i=1}^{4}{\bf H}^{[R,i]}\left({\bf v}^{[i]}_1
s_{i}+{\bf v}^{[i]}_2 s_{i}\right)+{\bf n}^{[R]},
\end{eqnarray}
where ${\bf v}^{[1]}_2$=${\bf v}^{[4]}_2$=${\bf 0}_{M \times 1}$.

\begin{figure}
\centering
\includegraphics[width=3.3in]{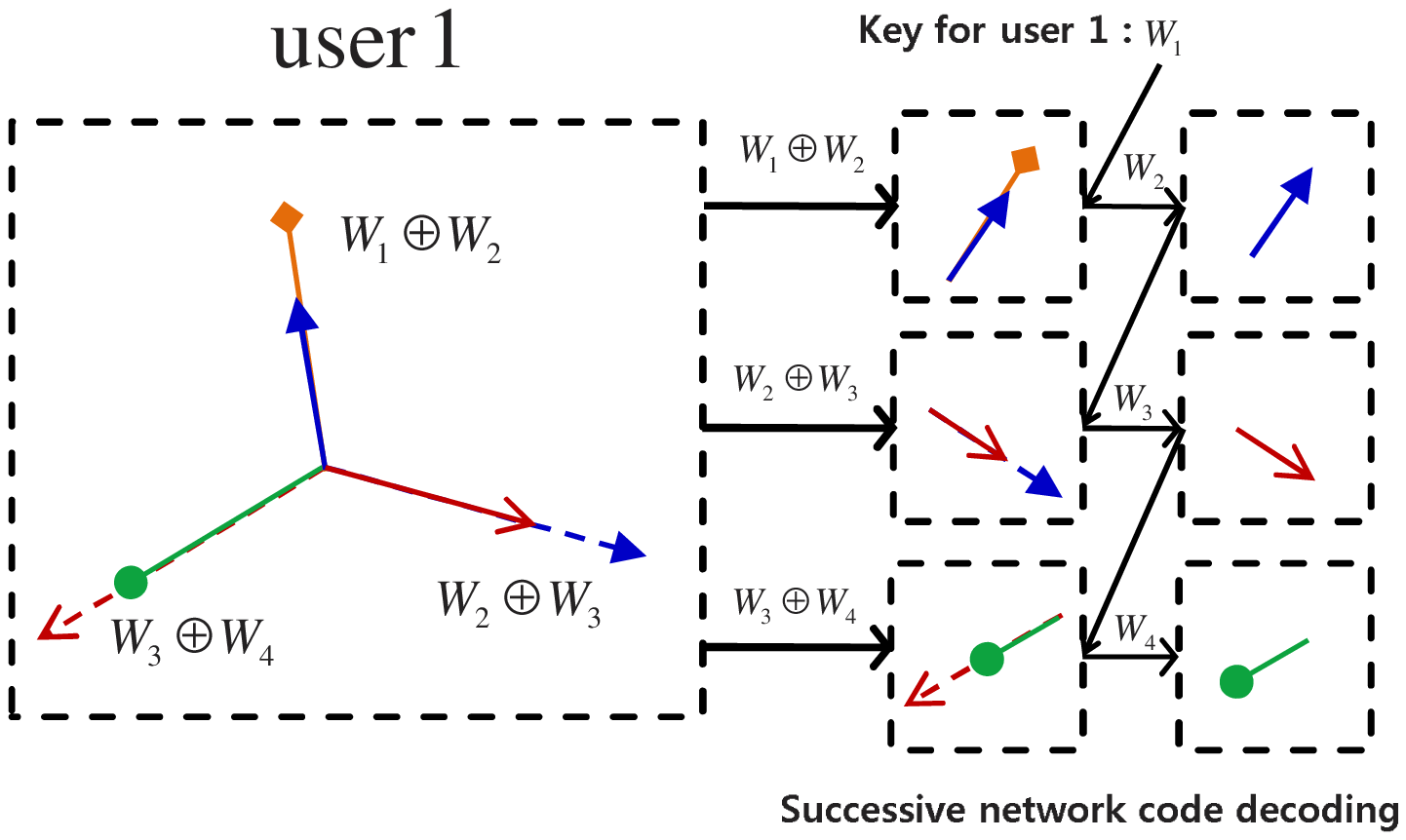}
\caption{Zero-forcing with successive network code decoding
example of user $1$ when $K=4$.} \label{fig_sim}
\end{figure}

The proposed precoding strategy is aiming at obtaining three
network coding messages with chain property, $W^{[1]}\oplus
W^{[2]}, W^{[2]}\oplus W^{[3]},W^{[3]}\oplus W^{[4]}$ at the
relay. Therefore, to accomplish this, precoding vectors are
constructed by exploiting signal space alignment for network
coding scheme \cite{namyoon1}. Fig. 2 illustrates the conceptual
figure of signal space alignment for an encryption message when
$K=4$. To decode $K-1=3$ network coded messages, $W^{[1]}\oplus
W^{[2]}, W^{[2]}\oplus W^{[3]},W^{[3]}\oplus W^{[4]}$ at the
relay, all users carefully chooses the precoding vectors in order
to satisfy the conditions of signal space alignment for an
encryption message. These are given by:
\begin{eqnarray}
\textrm{span}({\bf H}^{[R,1]}{\bf v}^{[1]}_1) &\doteq&
\textrm{span}({\bf
H}^{[R,2]}{\bf v}^{[2]}_1), \nonumber \\
\textrm{span}({\bf H}^{[R,2]}{\bf
v}^{[2]}_2) &\doteq& \textrm{span}({\bf H}^{[R,3]}{\bf v}^{[3]}_1), \nonumber \\
\textrm{span}({\bf H}^{[R,3]}{\bf v}^{[3]}_2)&\doteq&
\textrm{span}({\bf H}^{[R,4]}{\bf v}^{[4]}_1),
\end{eqnarray}
where $\textrm{span}({\bf A})\doteq \textrm{span}({\bf B})$
indicates that the column space of ${\bf A}$ and ${\bf B}$ are
identical. The transmit beamforming vectors satisfying the
conditions in (11) can be obtained by solving the
\emph{generalized eigenvalue problem} as
\begin{eqnarray}
{\bf v}^{[1]}_1={\alpha}^{[1]}_1{\bf f}_1, \quad {\bf
v}^{[2]}_1={\alpha}^{[2]}_1{\bf f}_1 \nonumber \\
{\bf v}^{[2]}_2={\alpha}^{[2]}_2{\bf f}_2, \quad {\bf
v}^{[3]}_1={\alpha}^{[3]}_1{\bf f}_2 \nonumber \\
{\bf v}^{[3]}_2={\alpha}^{[3]}_2{\bf f}_3, \quad {\bf
v}^{[4]}_1={\alpha}^{[4]}_1{\bf f}_3,
\end{eqnarray}
where ${\bf f}_1$, ${\bf f}_2$, and ${\bf f}_3$ are eigenvectors
that correspond to the maximum eigenvalues of matrices
$\left({{\bf H}^{[R,1]}}^{-1}{\bf H}^{[R,2]}\right)$, $\left({{\bf
H}^{[R,2]}}^{-1}{\bf H}^{[R,3]}\right)$, and $\left({{\bf
H}^{[R,3]}}^{-1}{\bf H}^{[R,4]}\right)$, respectively, and
${\alpha}^{[1]}_1$, ${\alpha}^{[2]}_1$, ${\alpha}^{[2]}_2$,
${\alpha}^{[3]}_1$,  ${\alpha}^{[3]}_2$ and ${\alpha}^{[4]}_1$ are
power normalization coefficients. These coefficients are
determined to satisfy the following conditions as
\begin{eqnarray}
{\alpha^{[1]}_1}^2\|{\bf H}^{[R,1]}{\bf f}_1\|^2 &=&
{\alpha^{[2]}_1}^2\|{\bf H}^{[R,2]}{\bf f}_1\|^2 , \label{eq:normal1} \nonumber \\
{\alpha^{[2]}_2}^2\|{\bf H}^{[R,2]}{\bf f}_2\|^2 &=&
{\alpha^{[3]}_1}^2\|{\bf H}^{[R,3]}{\bf f}_2\|^2 , \label{eq:normal2} \nonumber \\
{\alpha^{[3]}_2}^2\|{\bf H}^{[R,3]}{\bf f}_3\|^2 &=&
{\alpha^{[4]}_1}^2\|{\bf H}^{[R,4]}{\bf f}_3\|^2 , \label{eq:normal3} \nonumber \\
{\alpha^{[i]}_1}^2+{\alpha^{[i]}_2}^2 &\leq& \textsf{SNR},
~~~\forall i,
\end{eqnarray}
where $\alpha^{[1]}_2=\alpha^{[4]}_2=0$. For example, if $\|{\bf
H}^{[R,1]}{\bf f}_1\|^2 < \|{\bf H}^{[R,2]}{\bf f}_1\|^2$, we
design ${\alpha^{[1]}_1}^2=\textsf{SNR}/2$. By doing so,
${\alpha^{[2]}_1}^2 < \textsf{SNR}/2$. In addition, if $\|{\bf
H}^{[R,2]}{\bf f}_2\|^2 < \|{\bf H}^{[R,3]}{\bf f}_3\|^2$, we also
design ${\alpha^{[2]}_2}^2=\textsf{SNR}/2$. Therefore, user 2 can
satisfy the transmit power constraint,
${\alpha^{[2]}_1}^2+{\alpha^{[2]}_2}^2 \leq \textsf{SNR}$. In a
similar way, all power normalizing coefficients can be calculated
so that the output magnitude of two vectors on the same signal
dimension of the relay are the same while satisfying the transmit
power constraint of all users.

As the result, the received signal at the relay in (10) becomes
\begin{eqnarray}
{\bf y}^{[R]}&=&\left[%
\begin{array}{ccc}
 {\bf u}_{1} & {\bf u}_{2} & {\bf u}_{3} \\
\end{array}%
\right]\left[%
\begin{array}{c}
  {s_1}+{s_2} \\
  {s_2}+{s_3} \\
  {s_{3}}+{s_{4}} \\
\end{array}%
\right]+{\bf n}^{[R]},\\ \nonumber &=&{\bf U}^{[R]}{\bf
s}^{[R]}+{\bf n}^{[R]},
\end{eqnarray}
where ${\bf u}_1={\bf H}^{[R,1]}{\bf v}^{[1]}_1={\bf
H}^{[R,2]}{\bf v}^{[2]}_1$, ${\bf u}_2={\bf H}^{[R,2]}{\bf
v}^{[2]}_2={\bf H}^{[R,3]}{\bf v}^{[3]}_1$, and ${\bf u}_3={\bf
H}^{[R,3]}{\bf v}^{[3]}_2={\bf H}^{[R,4]}{\bf v}^{[4]}_1$. In
(14), ${\bf U}^{[R]}$ denotes the effective channel matrix. At
this point, it is necessary to check the decodability of ${\bf
s}^{[R]}$ at the receiver of the relay. The decodability is simply
proved by showing that $\mathrm{Pr}\left[\det({\bf
U}^{[R]})=0\right]=0$. Recall that ${\bf u}_{i}$, $i=\{1,2,3\}$,
is one of the $M$ intersection basis vectors between ${\bf
H}^{[R,i]}$ and ${\bf H}^{[R,i+1]}$. In addition, we define a
subspace, $V_i^{c}$, consisted by basis vectors $\left\{{\bf
u}_{1},{\bf u}_{2},{\bf u}_{3}\right\}-\left\{{\bf
u}_{i}\right\}$. As the entries of ${\bf H}^{[R,i]}$ are generated
by a continuous distribution, the probability that a basis vector
in the intersection of any two channel matrices lies in the other
intersection subspace spanned by the other two channel matrices is
zero, i.e., $\mathrm{Pr}\left[{\bf u}_{i} \subset V_i^c\right]=0$,
$\forall i$. Consequently, this gives
\begin{eqnarray}
\mathrm{Pr}\left[\det({\bf U}^{[R]})=0\right]\leq
\sum_{i=1}^{3}\mathrm{Pr}\left[{\bf u}_{i} \subset
V_i^{c}\right]=0.
\end{eqnarray}
Therefore, ${\bf s}^{[R]}$ can be obtained by eliminating the
inter-signal space interference using a zero-forcing decoder,
which is
\begin{eqnarray}
{\bf s}^{[R]}={\bf U}^{[R]^{-1}}{\bf y}^{[R]}+{\bf
U}^{[R]^{-1}}{\bf n}^{[R]}.
\end{eqnarray}
The three network coded messages
$\hat{W}^{[R]}_{\pi(1,2)}=W_1\oplus W_{2}$,
$\hat{W}^{[R]}_{\pi(2,3)}=W_2\oplus W_{3}$, and
$\hat{W}^{[R]}_{\pi(3,4)}=W_3\oplus W_{4}$ are then obtained by
applying the physical-layer network coding (PNC)
modulation-demodulation mapping principle \cite{Zhang1} into each
symbol $s^{[R]}_{i}$ via a signal dimension.

During the BC phase (the second time slot), the relay broadcasts
three encrypted information $\hat{W}^{[R]}_{\pi(1,2)}=W_1\oplus
W_{2}$, $\hat{W}^{[R]}_{\pi(2,3)}=W_2\oplus W_{3}$, and
$\hat{W}^{[R]}_{\pi(3,4)}=W_3\oplus W_{4}$ to all users using $3$
encoded symbols $\left[q^{[R]}_{1},q^{[R]}_{2},q^{[R]}_{3}\right]$
along beamforming vectors $\left[{\bf v}^{[R]}_{1},{\bf
v}^{[R]}_{2},{\bf v}^{[R]}_{3}\right]$. This is denoted as
follows:
\begin{equation}
{\bf x}^{[R]}=\sum_{i=1}^{3}{\bf v}^{[R]}_{i}{q^{[R]}_{i}}, \quad
i=1,2,3.
\end{equation}
The received signal at user $i$ can be expressed as
\begin{eqnarray}
{\bf y}^{[i]}&=&{\bf H}^{[i,R]}{\bf x}^{[R]}+{\bf n}^{[i]}, \nonumber\\
&=& {\bf H}^{[i,R]}\sum_{i=1}^{3}{\bf
v}^{[R]}_{i}{q^{[R]}_{i}}+{\bf n}^{[i]}, \nonumber\\
&=& {\bf Q}^{[i,R]}{\bf q}^{[R]}+{\bf n}^{[i]},
\end{eqnarray}
where ${\bf Q}^{[i,R]}$ denotes the effective channel from the
relay to user $i$, and ${\bf
q}^{[R]}=\left[q^{[R]}_{1},q^{[R]}_{2},q^{[R]}_{3}\right]^{T}$. In
this case, to detect the symbol ${\bf q}^{[R]}$ at user $i$, the
effective channel, ${\bf Q}^{[i,R]}$ should be invertible. This
condition is guaranteed if the precoding matrix ${\bf V}^{[R]}=
\left[{\bf v}^{[R]}_{1},{\bf v}^{[R]}_{2},{\bf
v}^{[R]}_{3}\right]$ has the rank of $3$ due to the fact that the
rank of the product of two square matrices ${\bf A}$ and ${\bf B}$
cannot exceed the smallest rank of the multiplicand matrices. In
other words, all effective channels ${\bf Q}^{[i,R]}$,
$i=\{1,2,3\}$ have full rank if ${\bf V}^{[R]}$ has the $3$ rank,
as all downlink channels from the relay to the users almost likely
have the rank of $3$. Therefore, the conclusion is that all users
are able to detect ${\bf q}^{[R]}$ by zero-forcing detection,
which nulls out the inter-signal space interference. For user $i$,
the detected signal is
\begin{eqnarray}
{\bf q}^{[R]}={\bf Q}^{[i,R]^{-1}}{\bf y}^{[i]}+{\bf
Q}^{[i,R]^{-1}}{\bf n}^{[i]}.
\end{eqnarray}

After detection ${\bf q}^{[R]}$, each user decodes the messages
from the other users by resolving the encrypted message,
$W^{[R]}_{\pi(i,i+1)}=W_i\oplus W_{i+1}$, $i=\{1,2,3\}$, which is
contained in ${\bf q}^{[R]}$. The decryption procedure is
performed by successive network code decoding scheme shown in Fig.
3. Considering the receiver of user 1, here, user 1 wants to
decode $\hat{W}_2, \hat{W}_3$, and $\hat{W}_4$, reliable which are
the messages coming from user 2, user 3, and user $4$,
respectively. By exploiting self-information $W_1$ as a key, user
1 first extracts $\hat{W}_2$ from message ${W}^{[R]}_{1}=W_1\oplus
W_2$ contained in ${q}^{[R]}_1$, as follows:
\begin{eqnarray}
 \hat{W}_{2}= W^{[R]}_{1} \oplus W_1 = (W_1\oplus W_2) \oplus W_1.
\end{eqnarray}
Subsequently, user 1 successively decodes $\hat{W}_3$ from message
$W^{[R]}_{2}=(W_2\oplus W_3)$, which is contained in ${q}^{[R]}_2$
using message $W_2$ as another key; this is decoded message from
the previous step. In this consecutive approach, user 1 untangles
the network code chain successively. Eventually, user 1 is able to
obtain all messages $\hat{W}_2, \hat{W}_3$, and $\hat{W}_4$ from
the other users.

In the same manner, other all users can resolve the encrypted
message $W^{[R]}_{\pi(i,i+1)}=W_i\oplus W_{i+1}$, $i=\{1,2,3\}$,.
Therefore, $\frac{K}{2}=\frac{4}{2}=2$ degrees of freedom can be
achieved on the MIMO Gaussian $4$-way relay channel. In other
word, \textit{4 users can exchange information each other within
two time slots on the MIMO Gaussian $4$-way relay channel if all
nodes have $M=K-1=3$ antennas}.

Therefore, the achievable degrees of freedom of the MIMO Gaussian
$4$-way relay channel according to two schemes is summarized as
shown in TABLE \ref{Table1}.

%

\begin{table}[tp]%
\caption {The sum of achievable degrees of freedom when $K=4$.}
\centering %
\begin{tabular}{|c|c|c|}
  \hline
  Scheme & Sum of degrees of freedom ($\eta$)  & Number of antennas ($M$)\\
  \hline\hline
  TDMA & $\frac{K-1}{2}=\frac{3}{2}$ &  $K-1=3$\\ \hline
  Proposed & $\frac{K}{2}=2$ & $K-1=3$ \\
  \hline
\end{tabular} \label{Table1}
\end{table}
%
%
%
\emph{Remark 1}: In the Gaussian single-input-single-output (SISO)
$K$-way relay channel, $\frac{K}{K-1}$ degrees of freedom can be
achieved if the channel coefficients are time/frequency variant.
In this case, the precoding vectors for the signal space alignment
for an encryption message are designed over the $K-1$
time/frequency slots, similar to the method used in
\cite{Cadambe}.

\emph{Remark 2}: In the BC phase, the relay transmits the
encrypted massages with the network code chain structure. Assuming
that an eavesdropper having $M$ antennas wants to decode the
messages during the BC phase, the question arises as to how many
messages the eavesdropper can reliably decode. The proposed coding
strategy only allows the users who participate in the message
exchange via relay to decipher the messages, as the users who take
part have the key to crack the network code chain. Therefore, due
to the absence of knowledge regarding a certain key, the
eavesdropper cannot decode any message. It indicates that the
proposed scheme is robust in terms of the message security.

%
%

\section{Conclusions} \label{sec:Conclusions}
In the MIMO Gaussian $K$-way relay channel, we proposed an
encoding and decoding strategies inspired from cryptography. It
involves \textit{signal space alignment for an encryption message}
and \textit{zero forcing combined with successive network code
decoding}. Using this encoding decoding scheme, it was shown that
the proposed scheme can achieve the higher degrees of freedom as
compared with that of TDMA scheme.

\end{document}